\documentclass[prX,showpacs,superscriptaddress,twocolumn ]{revtex4}

\usepackage{graphicx}
\usepackage{amsmath}
\usepackage{amsfonts}
\usepackage{amssymb}
\usepackage{epsf}
\usepackage{hyperref}

\begin{document}
\title{Exploring classically chaotic potentials with a matter wave quantum probe}
\author{G. L. Gattobigio}
\affiliation{Laboratoire de Collisions Agr\'egats R\'eactivit\'e,
CNRS UMR 5589, IRSAMC, Universit\'e de Toulouse (UPS), 118 Route de
Narbonne, 31062 Toulouse CEDEX 4, France}
\affiliation{Laboratoire Kastler Brossel, Ecole Normale
Sup\'erieure, 24 rue Lhomond, 75005 Paris, France}
\author{A. Couvert}
\affiliation{Laboratoire Kastler Brossel, Ecole Normale
Sup\'erieure, 24 rue Lhomond, 75005 Paris, France}
\author{B. Georgeot}
 \affiliation{Laboratoire de Physique Th\'eorique (IRSAMC), Universit\'e de Toulouse 
(UPS), 31062 Toulouse, France} 
\affiliation{CNRS, LPT UMR5152 (IRSAMC), 31062 Toulouse, France}

\author{D. Gu\'ery-Odelin}
\affiliation{Laboratoire de Collisions Agr\'egats R\'eactivit\'e,
CNRS UMR 5589, IRSAMC, Universit\'e de Toulouse (UPS), 118 Route de
Narbonne, 31062 Toulouse CEDEX 4, France}

 \date{\today}

\begin{abstract}
We study an experimental setup in which a quantum probe, provided by a quasi-monomode guided atom laser, interacts with a static localized attractive potential whose characteristic parameters are tunable. In this system, classical mechanics predicts a transition from regular to chaotic behavior as a result of the coupling between the different degrees of freedom. Our experimental results display a clear signature of this transition. On the basis of extensive numerical simulations, we discuss the quantum versus classical physics predictions in this context. This system opens new possibilities for investigating quantum scattering, provides a new testing ground for classical and quantum chaos and enables to revisit the quantum-classical correspondence. 
\end{abstract}

\maketitle

The realization and the manipulation of Bose-Einstein condensates (BECs) have led to many experiments on atom-atom interactions in the low energy regime: low partial waves elastic scattering has been analyzed to deduce accurate determination of collisional properties\cite{BLK04,TKJ04}, the possibility of tailoring the scattering length using either magnetic \cite{IAS98} or optical Feshbach resonances \cite{OFB} has been demonstrated, and the role of the dimensionality on atom-atom scattering processes through very anisotropic confinement has been studied \cite{PeS01,BMO03}.

In all these examples, the BEC is used to reveal the microscopic atom-atom interactions properties. BECs have also been placed into disordered \cite{BJZ08,DPH10} or non-commensurable bi-chromatic potentials \cite{Florence}.  
Such experiments do not study a single matter wave scattering event. This is to be contrasted with experiments where a BEC bounces on a mirror \cite{mirror}. However, in this case, the potential with which the BEC interacts is extremely simple. 

\begin{figure}[t]
\centerline{\includegraphics[width=7cm]{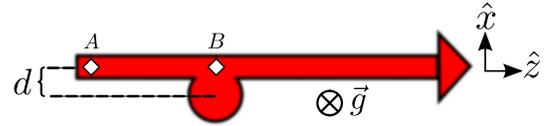}}
\caption{(color online). Sketch of the experimental setup. Atoms outcoupled from a BEC at point $A$ are guided by the horizontal beam towards point $B$. In the vicinity of $B$, atoms also experience the attractive force exerted by the LAP (see text) whose center is at a distance $d$ from the guide axis.} \label{fig1}
\end{figure}

In this article, we demonstrate the use 
of a guided matter-wave \cite{GRG06,CJK08,GCJ09,KWE10} 
to study, in a confined environment, the interaction of a matter wave with a static tunable localized attractive potential (LAP) associated with a complex dynamics. The transverse confinement of the guide can be well approximated by an harmonic confinement. 
The LAP superimposed to the guide breaks the harmonic character of the potential experienced by the atoms and couples the longitudinal and transverse degrees of freedom (see Fig.~\ref{fig1}). We study this scattering problem theoretically from the weak to the strong coupling regimes for which the system exhibits classical chaos. This can be done by tuning the LAP power (strength) and position $d$. Finally, we show that these different regimes can be clearly identified in our experimental results. 

Our experiment provides a direct access to the modulus squared of the wave function through absorption imaging. This is to be contrasted with experiments carried out with microwave in chaotic structure in which the measurement is performed at a well-defined position \cite{LVP00}. In mesoscopic physics, experiments exploit the conductance measurements to infer properties of the chaotic structure \cite{sols,JBS90,Bee97,heller}. However, in these systems, the chaotic dynamics effects cannot be singled out from disorder and interactions effects. 
In contrast with pioneering experiments using cold atoms to investigate classical chaotic behavior in static optical billiards \cite{FKC01,MHC01}, in our experiment we prepared atoms in a nearly pure quantum state and we analyzed a single event rather than relying on average observables. 
Other studies on quantum chaos have been performed with cold atoms in the different context of 1D time-dependent lattice potentials \cite{MRB95,SRD02,RAV06}.

The experimental setup that delivers guided matter waves has been described in detail elsewhere \cite{GCJ09}. A rubidium-87 BEC is produced in a crossed dipole trap using two focused laser beams with a wavelength of 1070 nm and waists $w_{\rm G}\approx$40 $\mu$m. Atoms are prepared in the $|F=1,m_F=0\rangle$ internal state through a spin distillation process implemented during the evaporative cooling stage \cite{CJK08}. The quantum probe is realized by magnetic or all-optical outcoupling from the trap (located at $A$ in Fig.~\ref{fig1}) \cite{GCJ09}. Atom-atom interactions turn out to be negligible in the guided atom lasers produced in this manner \cite{GCJ09}. Experimentally, the propagating matter wave is characterized by three quantities: the atom flux (a few 10$^5$ atoms/s), the mean velocity $\bar{v}$ (tunable from 10 to 30 mm.s$^{-1}$) and the population of the transverse modes.  
Experimental parameters are chosen to have the mean excitation number $\langle n \rangle$ below one, which corresponds to a ground state population larger than 50\% \cite{GCJ09}. Such a probe therefore explores the phase space with a transverse resolution close to $\hbar^2$.

The horizontal beam used for the dipole trap provides the guiding potential which can be approximated by a harmonic transverse confinement with a measured frequency $\omega_\perp / 2\pi$ ranging from 170 Hz to 200 Hz (guide power, $P_{\rm G}\simeq$ 100-150 mW). The LAP originates from the same laser source as for the crossed dipole trap and is slightly detuned to wash out interferences. It has a waist of $w_{\rm LAP}\approx 60$ $\mu$m and an adjustable power, $P_{\rm LAP}$, from 0 to 2W. It intersects the guide beam at a distance $AB=700$ $\mu$m from the trap ($A$) along the propagation direction (see Fig.~\ref{fig1}). 

\begin{figure}[t]
\centerline{\includegraphics[width=\columnwidth]{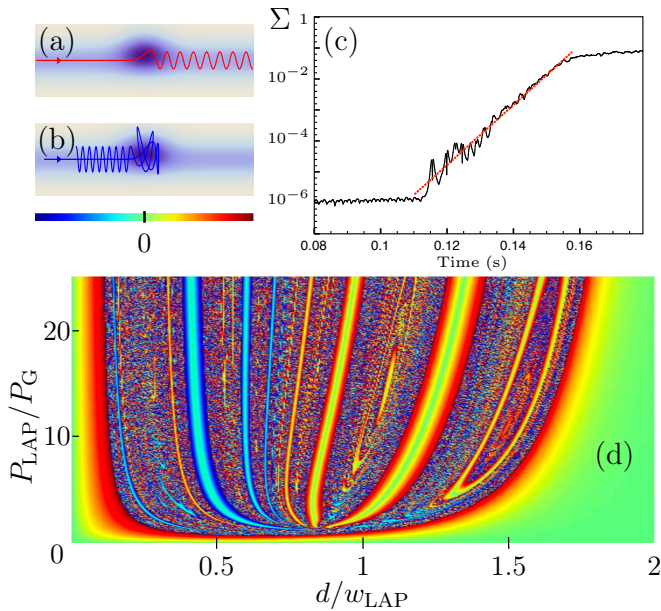}}
\caption{(color online). (a) Typical trajectory in the weak coupling regime (see text). The interaction with the LAP yields a downstream propagation with a transverse oscillation. (b) Example of trajectory in the strong coupling regime. The complex dynamics may lead to the reflexion of the particle. In both (a) and (b) the LAP potential is represented by the blue shaded area. (c) Example of exponential separation $\Sigma$ of nearby trajectories in phase space revealing the chaotic nature of the potential. The red dotted curve is the best exponential fit. Average over $100$ couples of trajectories close to the central one is taken. (d) Amplitude of the transverse excitation as a function of the ratio $P_{\rm LAP}/P_{\rm G}$ and the displacement $d$ of the LAP: transmitted (red colors) and reflected (blue colors) particles. A chaotic zone with a complex dynamics can be clearly identified.} \label{fig2}
\end{figure}

The interaction of the quantum probe with a well-centered LAP ($d=0$) has been investigated theoretically in \cite{NJP10}. To analyze the experimental results of the scattering problem for $d\neq 0$ it is instructive, as a first guideline, to use classical mechanics. 

Consider that atoms are classical point-like particles delivered at the trap location ($A$) and at the bottom of the guide (see Fig.~\ref{fig1}). As a result of the longitudinal curvature of the guide beam, atoms are accelerated towards point $B$. Close to $B$, atoms feel the combination of the guide and the attractive potential. When the LAP is well centered ($d=0$), it does not generate a transverse oscillation downstream. However, when the transverse symmetry is broken ($d\neq 0$), the LAP has a dramatic effect. In the weak coupling regime, 
the beam goes downstream and exhibits a transverse dipole oscillation (Fig.~\ref{fig2}a).  A concomitant reduction of the longitudinal velocity occurs due to energy conservation. This output channel is stable against small changes of $d/w_{\rm LAP}$ and $P_{\rm LAP}/P_{\rm G}$. In the strong coupling regime, the output channel can be either upstream  (Fig.~\ref{fig2}b) or downstream depending on the different parameters of the LAP (strength and displacement) and turns out to be extremely sensitive to small parameters changes (Fig.~\ref{fig2}d). 

 The complex trajectories that result from the strong coupling between longitudinal and transverse degrees of freedom illustrate the chaotic nature of the potential in this regime (see Fig.~\ref{fig2}c) where the Lyapunov exponents can be shown to be positive and the distribution of trapping times is found to exhibit an exponential decay (data not shown). Figure~\ref{fig2}d summarizes the variation of the asymptotic transverse amplitude as a function of the ratio $P_{\rm LAP}/P_{\rm G}$ and the displacement $d$ of the LAP. It exhibits a region with strong variations of output channels with parameters, that we identify as the chaotic zone. To further characterize this zone, we have verified that the set of boundaries between left and right output channels as a function of $d$ displays a fractal structure with a fractal dimension $\xi$. For a wide variety of fixed ratio $P_{\rm LAP}/P_{\rm G}$ in the chaotic zone, we find an exponent $\xi \approx 0.9$ revealing a strong fractality of the basin boundaries, usually associated with chaotic scattering \cite{Ott}.

However, the incident guided atom laser is not a classical probe since it is in the ground state of the transverse confinement \cite{GCJ09}. In addition, the potential experienced by the atoms is not harmonic and the dynamics cannot thus be exactly mapped onto the classical one. 

\begin{figure}[t]
\centerline{\includegraphics[width=8cm]{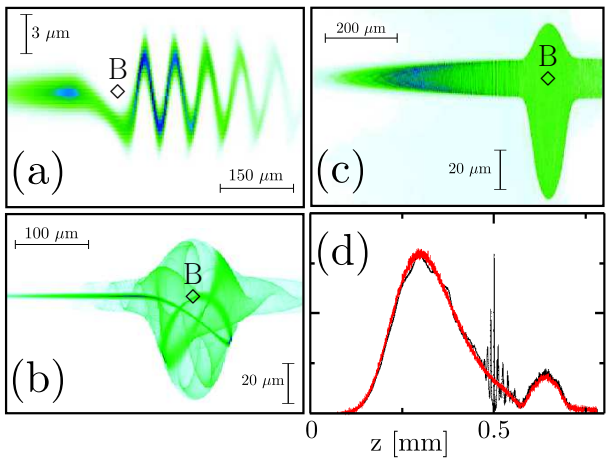}}
\caption{(Color online). (a) (b) (c) Color density plots of the wave
function from 2D numerical simulations. Color scale is not linear. Point $B$ is the same as in Fig.~\ref{fig1}.
Wave packets are prepared in the ground state of the initial trap, and released in the guide at time $t=0$ ($w_{\rm G}=45$ $\mu$m and $\omega_\perp= 2\pi \times 203$ Hz) at 700 $\mu$m from the LAP ($w_{\rm LAP}=41$ $\mu$m). The incident mean velocity of the wave packet is $\bar{v}\sim$ 10 mm.s$^{-1}$ close to $B$. (a) $d=2$ $\mu$m ($d/w_{\rm LAP}=0.05$), $P_{\rm LAP}=0.5$ W ($P_{\rm LAP}/P_{\rm G}=2.78$),  propagation time $t=120$ ms, (b):
$d=15$ $\mu$m ($d/w_{\rm LAP}=0.37$) and $P_{\rm LAP}=1$W ($P_{\rm LAP}/P_{\rm G}=5.56$), $t=120$ ms,
(c) $d=20$ $\mu$m ($d/w_{\rm LAP}=0.5$), $P_{\rm LAP}=1$ W ($P_{\rm LAP}/P_{\rm G}=5.56$), $t=160 $ ms, (d) projected distribution (c) on the $z$ axis
of the wave function $\int {\rm d}x |\psi(x,z)|^2$ (black) and the initially identical
classical distribution (red/gray); classical and quantum
curves are identical for cases (a) and (b) (data not shown).}
\label{fig4}
\end{figure}

To explore the quantum-classical correspondence, we have developed a full 2D numerical simulation of the wave packet
that represents the incident elongated matter wave. The extremely large anisotropy of the system with a
longitudinal length of a few mm and a transverse size initially of a few hundreds of nm and the long interaction time inherent to the complex dynamics makes this kind of calculation memory size and time consuming, and even unrealistic in 3D with the currently available supercomputers. We have thus compared the dynamics of 2D quantum
wavepackets with simulations of classical probability densities of same initial position and momentum
 distributions. Parameters are chosen in close similarity with the experimental conditions.
In the weak coupling regime, the scattering is regular, the quantum wave packet keeps its structure and performs coherent oscillations downward the LAP (Fig.~\ref{fig4}a). In this regime, classical and quantum predictions are indistinguishable. In the strong coupling regime for which the classical counterpart is chaotic, the wave packet loses its filamentary structure.
In this regime, the numerically simulated wavepacket follows the central trajectory for short time (see
Fig.~\ref{fig4}b). For larger time, it loses its structure and classical
and quantum dynamics can become different (see Fig.~\ref{fig4}c and
Fig.~\ref{fig4}d). The quantum and classical results differ significantly when the wave function overlaps with itself in the LAP region and therefore exhibits some interference patterns (see Fig.~\ref{fig4}d). 

\begin{figure}[t]
\centerline{\includegraphics[width=8cm]{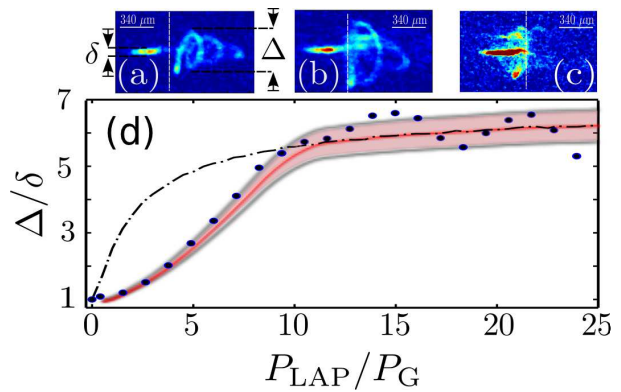}}
\caption{(color online). Interaction between a guided atom laser and a LAP (experimental results). Absorption image (a) is taken in the weak coupling regime and, (b) and (c) in the strong coupling regime after a 15 ms time-of-flight. (d) Measurement of the displacement $d$ in the weak coupling regime: the ratio $\Delta/\delta$ of the transverse size before and after the interaction is plotted vs $P_{\rm LAP}/P_{\rm G}$ ($\bar{v} \simeq 25$ mm.s$^{-1}$, and $\omega_\perp/2\pi \simeq 200$ Hz). The dot-dashed curve corresponds to 3D (classical) simulation with $d=4$ $\mu$m ($d/w_{\rm LAP}\approx 0.07$). Most experimental points (disks) are inside the pink area delimited by the simulated curves (gray lines) for 3.7 and 4.3 $\mu$m.} \label{fig3}
\end{figure}

The scattering problem depends on many parameters: those that characterize the incident wave (velocity, number of occupied transverse modes, longitudinal coherence length, ...) and those that concern the confinement (powers, waists and distance $d$). Experimentally, we have investigated this large parameter space by varying the ratio $P_{\rm LAP}/P_{\rm G}$ which  in turn slightly modifies the displacement $d$ (see below). 
 The experiment duration, $\tau \sim 300$ ms, is chosen so that typically half the guided atom laser has interacted with the LAP in the weak coupling regime. This duration was chosen in order to observe the asymptotic regime in the weak coupling regime and the complex dynamics associated with the strong coupling regime. Absorption images are taken after time-of-flight once the confinement potentials are switched off. In this way, the ratio between the final, $\Delta$, and initial, $\delta$, transverse sizes can readily be  extracted.  Examples of experimental results are shown in Fig.~\ref{fig3}. The two regimes (regular and chaotic behaviors) are clearly observed when the power is increased (see Fig.~\ref{fig3}a and, Fig.~\ref{fig3}b and c respectively). The transverse oscillation observed after the LAP in the weak interacting regime corresponds to a coherent transverse dipole oscillation (Fig.~\ref{fig3}a) in close similarity with the corresponding quantum simulation (Fig.~\ref{fig4}a). The apparent decrease of the transverse size of the helix motion is due to the guide anisotropy and the projection of the trajectory along the imaging axis. 

For given parameters, the displacement $d$ of the LAP is reproducible from shot to shot. However, $d$ varies slightly with the LAP power. This latter effect is due to thermal effect generated by the RF power in the acousto-optic modulator that control the LAP power. In the parameter space of Fig.~\ref{fig2}d, the experimental results thus do not follow a vertical line. This is clearly illustrated in Fig~\ref{fig3}d where we have represented $\Delta/\delta$, in the weak coupling regime, for both the experimental points, and 3D (classical) simulations that assume a fixed displacement $d=4$ $\mu$m. Here, we use the fact that classical and quantum predictions coincide in this regime. The observed saturation for  $P_{\rm LAP} > 10 P_{\rm G}$ suggests that in this parameter regime $d$ is approximately constant $d\simeq 4$ $\mu$m $\pm$ 0.3 $\mu$m. For $P_{\rm LAP} < 10 P_{\rm G}$, the initial curvature is opposite to the simulated one revealing a change of $d$ as $P_{\rm LAP}$ is increased. Using a linear increase of $d$ up to  $P_{\rm LAP} =10 P_{\rm G}$ towards the value $d=4$ $\mu$m and a constant value afterwards gives a very good agreement with the experimental data (see Fig.~\ref{fig3}d). 

Interestingly enough, the experiments turn out to be robust with respect to the velocity dispersion of the incoming matter wave. This effect, confirmed by numerical simulations, originates from the fact that the larger available energy for larger incident velocity is compensated for by the smaller interaction time with the LAP.

Experimentally, we observe two types of images in the chaotic regime: those for which the trajectory can be well-identified (Fig.~\ref{fig3}b) and those with a more diffuse cloud (Fig.~\ref{fig3}c). Numerical simulations (see Fig.\ref{fig4}b) suggest that the former can be attributed to situations for which the chaotic dynamics did not have enough time to fully develop. The latter may result from either the ergodic mixture as in Fig.~\ref{fig4}c that occurs in the chaotic regime for sufficiently long time or situations in which the relatively high local atomic density favors elastic collisions and thus the possibility of exploring a larger phase space zone with various output channels.

\begin{figure}[t]
\centerline{
\includegraphics[width=8cm]{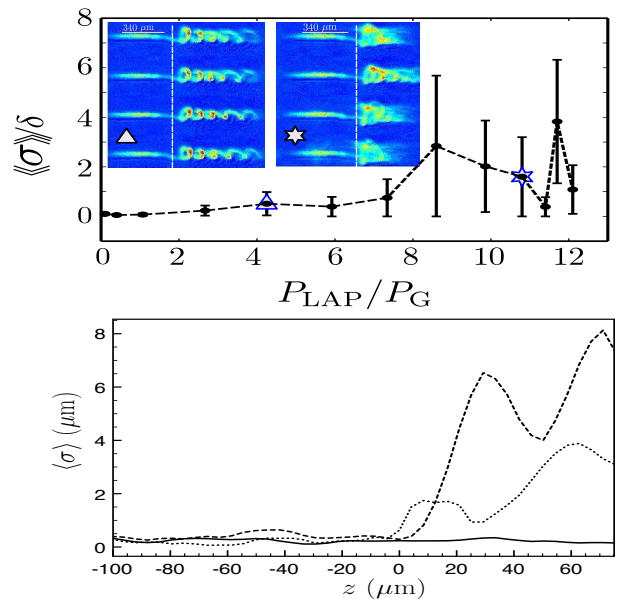}}
\caption{(color online) Experimental difference between quantum evolutions for different runs with same initial conditions (same parameters as in Fig.~\ref{fig3}). (Top) $\langle\langle \sigma \rangle\rangle$ (see text) vs $P_{\rm LAP}/P_{\rm G}$. The onset of the chaotic regime is clearly visible on both the large value of $\langle\langle \sigma \rangle\rangle$ and its error bar. Inset: Examples of these different runs in the weak (left) and strong (right) coupling regimes. (Bottom) $\langle \sigma(z) \rangle$ (see text) vs $z$. Solid line is $P_{\rm LAP}=0.06$ W ($P_{\rm LAP}/P_{\rm G}=0.6$), dotted line is $P_{\rm LAP}= 0.6$ W ($P_{\rm LAP}/P_{\rm G}=6$), dashed line is $P_{\rm LAP}= 1$ W ($P_{\rm LAP}/P_{\rm G}=10$). We observe the amplification of small initial deviations when the beam enters the chaotic zone ($z\sim z_{\rm B}\sim 0$) as in Fig.~\ref{fig2}c.} \label{fig5}
\end{figure}

In Fig.~\ref{fig5}, we assess the reproducibility of the experiment by using different runs with same initial conditions, where small initial differences are inevitably present. Examples of the different experimental images are shown in the top inset. We define the quantity $\sigma_{i,j}(z)$ as the difference of the mean transverse size of the wave packet at coordinate $z$ downward the LAP (located at $z_B$) for two different runs $i$ and $j$ taken in the same conditions. We note $\langle \sigma(z) \rangle = N_r^{-1}\sum_{i<j}\sigma_{i,j}(z)$ where $N_r$ is the number of runs ($N_r=5$ in our experiment) and $\langle\langle \sigma \rangle\rangle$ is $\langle \sigma(z) \rangle$ averaged over $z>z_{\rm B}$. In Fig.~\ref{fig5} (top), this latter quantity is represented vs  $P_{\rm LAP}/P_{\rm G}$. 
 When the system enters the chaotic (strong coupling) regime a large dispersion in the transverse size is observed as a result of the high sensitivity of the system to parameter changes and initial conditions.  The transition to chaos becomes visible for $P_{\rm LAP}/P_{\rm G}\sim 8$, compatible with values of $d \approx 7.5$ $\mu$m ($d/w_{\rm LAP}=0.125$) (see Fig.~\ref{fig2}d).

To make more visible the chaotic regime in the experiments, in Fig.~\ref{fig5} (bottom) we show 
the average distance $\langle \sigma(z) \rangle $. As long as the wavepacket moves along the $z$ coordinate, this coordinate gives the time evolution of the quantum wave function. Unavoidably small differences were present in the initial conditions. In the regular regime, these differences remain approximately of the same order of magnitude, while
in the chaotic regime they display a sharp exponential amplification when the wavepacket enters the chaotic potential. This is
reminiscent of what happens for classical trajectories (see Fig.~\ref{fig2}c). Strictly speaking, $\langle \sigma(z) \rangle $ does not give a direct access to the Lyapunov exponent since more runs would be required and we have only a projection of the density probability in position space (instead of the phase space trajectory in classical physics). However, according to our numerical studies, $\langle \sigma(z) \rangle $ in many cases can be used to give a lower bound to the Lyapunov exponent.

In conclusion, we have explored the effects of a LAP on a guided atom laser. A classical analysis enables to identify different regimes, from regular to chaotic and unstable dynamics. These different regimes are also visible in the quantum simulations as well as in the experiments. Differences between classical and quantum dynamics are especially noticeable in the non-regular regime.
This system enables to investigate both classical and quantum chaos in a well-controlled environment.
Two situations should be especially adapted to magnified quantum features: (i) the 2D geometry which favors interferences of the wavefunction with itself and (ii) situations for which LAP size is on the order of the transverse size of the initial wave function. In this latter regime quantum diffraction and/or tunneling effects should play an important role. The experimental exploration of these two regimes provides a natural continuation of our work. Another direction for future work includes the study of the nonlinear influence of atom-atom interaction and in particular matter wave solitons \cite{soliton,MGS08}.
 We think our approach paves the way for the use of guided matter waves to study highly non-equilibrium dynamics and to use such systems as quantum simulators for complex quantum scattering problems.

We thank R. Mathevet and T. Lahaye for useful comments.
We acknowledge financial support from the Agence Nationale de la Recherche (GALOP project), the R\'egion Midi-Pyr\'en\'ees, the university Paul Sabatier (OMASYC project) and the Institut Universitaire de France. We thank CalMip for the use of their supercomputers.

\end{document}